\documentclass[twocolumn,notitlepage,prb,color,superscriptaddress,psfig,showpacs,amsmath,amssymb,nobibnotes,longbibliography]{revtex4-2}

\usepackage[colorlinks=true,citecolor=blue]{hyperref}%
\usepackage{graphicx}
\usepackage{dcolumn}
\usepackage{bm}
\usepackage{verbatim}
\usepackage{epsf}
\usepackage{amssymb}
\usepackage{color}
\usepackage{float}
\usepackage{epstopdf}
\usepackage{natbib}
\usepackage{amsmath}
\usepackage{mathrsfs}
\usepackage{xspace}
\usepackage{placeins}

\usepackage{soul}
\setstcolor{red}

\newcommand{\MPS}{Mn$_{2}$P$_{2}$S$_{6}$\xspace}
\newcommand{\MNPS}{MnNiP$_{2}$S$_{6}$\xspace}
\newcommand{\NPS}{Ni$_{2}$P$_{2}$S$_{6}$\xspace}
\newcommand{\PS}{(Mn$_{1-x}$Ni$_x$)$_2$P$_2$S$_6$\xspace}

\newcommand{\HIIab}{$\textbf{H} \perp c$*\xspace}
\newcommand{\HIIc}{$\textbf{H} \parallel c$*\xspace}

\definecolor{darkgreen}{rgb}{0, 0.4, 0}

\begin{document}

\title{Magnetic anisotropy and low-energy spin dynamics in the van der Waals compounds \MPS and \MNPS}

\author{J.~J.~Abraham}
\thanks{These authors contributed equally to this work.}
\affiliation{Leibniz IFW Dresden, D-01069 Dresden, Germany}
\affiliation{Institute for Solid State and Materials Physics, TU Dresden, D-01062 Dresden, Germany}
\author{Y.~Senyk}
\thanks{These authors contributed equally to this work.}
\affiliation{Leibniz IFW Dresden, D-01069 Dresden, Germany}
\affiliation{Institute for Solid State and Materials Physics, TU Dresden, D-01062 Dresden, Germany}
\author{Y.~Shemerliuk}
\affiliation{Leibniz IFW Dresden, D-01069 Dresden, Germany}
\author{S.~Selter}
\affiliation{Leibniz IFW Dresden, D-01069 Dresden, Germany}
\affiliation{Institute for Solid State and Materials Physics, TU Dresden, D-01062 Dresden, Germany}
\author{S.~Aswartham}
\affiliation{Leibniz IFW Dresden, D-01069 Dresden, Germany}
\author{B.~B\"uchner}
\affiliation{Leibniz IFW Dresden, D-01069 Dresden, Germany}
\affiliation{Institute for Solid State and Materials Physics and W{\"u}rzburg-Dresden Cluster of Excellence ct.qmat, TU Dresden, D-01062 Dresden, Germany}
\author{V.~Kataev}
\affiliation{Leibniz IFW Dresden, D-01069 Dresden, Germany}
\author{A.~Alfonsov}
\affiliation{Leibniz IFW Dresden, D-01069 Dresden, Germany}
\affiliation{Institute for Solid State and Materials Physics and W{\"u}rzburg-Dresden Cluster of Excellence ct.qmat, TU Dresden, D-01062 Dresden, Germany}

\date{\today}

\begin{abstract}
	
	We report the detailed high-field and high-frequency electron spin resonance (HF-ESR) spectroscopic study of the single-crystalline van der Waals compounds \MPS and \MNPS. Analysis of magnetic excitations shows that in comparison to \MPS increasing the Ni content yields a larger magnon gap in the ordered state and a larger g-factor value and its anisotropy in the paramagnetic state. The studied compounds are found to be strongly anisotropic having each the unique ground state and type of magnetic order. Stronger deviation of the g-factor from the free electron value in the samples containing Ni suggests that the anisotropy of the exchange is an important contributor to the stabilization of a certain type of magnetic order with particular anisotropy. At temperatures above the magnetic order, we have analyzed the spin-spin correlations resulting in a development of slowly fluctuating short-range order. They are much stronger pronounced in \MNPS compared to \MPS. The enhanced spin fluctuations in \MNPS are attributed to the competition of different types of magnetic order. Finally, the analysis of the temperature dependent critical behavior of the magnon gaps below the ordering temperature in \MPS suggests that the character of the spin wave excitations in this compound undergoes a field induced crossover from a 3D-like towards 2D XY regime.
		
\end{abstract}

\maketitle

\section{Introduction}

	In the past recent years magnetic van der Waals (vdW) materials have become increasingly attractive for the fundamental investigations since they provide immense possibility to study intrinsic magnetism in low dimensional limit \cite{burch2018, huang2017, Cai2019}. The weak vdW forces hold together the atomic monolayers in vdW crystals, which results in a poor interlayer coupling, and therefore renders these materials intrinsically two dimensional. In addition to the fundamental research, these materials are very promising as potential candidates for next-generation spintronic devices \cite{khan2020, Li2019, scheunert2016, jimenez2020}. 
	
	Among the variety of magnetic vdW materials a particularly interesting subclass is represented by the antiferromagnetic (TM)$ _{2} $P$ _{2} $S$ _{6} $ tiophosphates (TM stands for a transition metal ion). Here the transition metal ions are arranged in a graphene-like layered honeycomb lattice \cite{Wang2018a}. The high flexibility of the choice of the TM ion enables to control the properties. Among the tiophosphates there are examples of superconductors \cite{Wang2018}, photodetectors and field effect transistors \cite{Jenjeti2018, Chu2017}. They also can be used for ion-exchange applications \cite{Joy1992}, catalytic activity \cite{Zhu2018}, etc. Therefore, a proper choice of TM, or of a mixture of magnetically inequivalent ions on the same crystallographic position, could lead to the possibility of engineering of a material with desired magnetic ground state, excitations and correlations. 
	
	In order to establish the connection between the choice of magnetic ion and the resulting ground state and correlations, we performed a detailed high-field and high-frequency electron spin resonance (HF-ESR) spectroscopic study on single crystals of the van der Waals compounds \MPS and \MNPS in a broad range of microwave frequencies and temperatures below and above the magnetic order. ESR spectroscopy is a powerful tool that can provide insights into spin-spin correlations, magnetic anisotropy and spin dynamics. This technique has shown to be very effective for exploration of the magnetic properties of vdW systems \cite{Okuda1986, joy1993, Lifshitz1982, Sibley1994, kobets2009, zeisner2019, wellm2018, zeisner2020, alahmed_2021, shen21, kavita2022, senyk2022}. Albeit resonance studies on \MPS were made by \citet{Okuda1986}, \citet{joy1993} and \citet{kobets2009}, a high-frequency ESR study exploring broad range of temperatures below and above magnetic order was not yet performed. The \MNPS compound is barely explored from the point of view of spin excitations from the magnetic ground state below the ordering temperature, and from the point of view of spin-spin correlations in the high temperature regime. 
	
	Investigating \MPS and \MNPS we have found difference in the types of magnetic order, anisotropies below the ordering temperature $T_N$, as well as the g-factors and their anisotropy above $T_N$ in these compounds. In fact, increasing the Ni content yields a larger magnon gap in the ordered state ($T<<T_N$) and a larger g-factor value and its anisotropy in the paramagnetic state ($T>>T_N$). At temperatures above the magnetic order, we have analyzed the spin-spin correlations resulting in a development of slowly fluctuating short-range order. They are much stronger pronounced in \MNPS compared to \MPS, which in our previous study has shown clear cut signatures of two-dimensional (2D) correlated spin dynamics \cite{senyk2022}. Therefore, enhanced spin fluctuations in \MNPS are attributed to the competition of different types of magnetic order. Finally, the analysis of the temperature dependent critical behavior of the magnon gaps below the ordering temperature in \MPS suggest that the character of the spin wave excitations in this compound undergoes a field induced crossover from a 3D-like towards 2D XY regime.

\section{Experimental Details}
\label{sec:meth}

	Crystal growth of \MPS and \MNPS samples investigated in this work was done using the chemical vapor transport technique with iodine as the transport agent. Details of their growth, crystallographic, compositional and static magnetic characterization are described in Refs. \cite{Selter21,shemerliuk2021}. Note, that the experimental value $x_{exp}$ in \PS for the nominal \MNPS compound is found to be $x_{exp} = 0.45$, considering an uncertainty of approximately 5\% \cite{shemerliuk2021}. Both materials exhibit a monoclinic crystal lattice system with a $ C2/m $ space group \cite{shemerliuk2021, Chica2021}. Each unit cell contains a [P$ _{2} $S$ _{6} $]$ ^{4-} $ cluster with S atoms occupying the vertices of TM octahedra and P-P dumbbells occupying the void of each metal honeycomb sublattice. The crystallographic $c$-axis makes an angle of 17$^{\circ}$ with the normal to the $ab$-plane \cite{Klingen1968}, which is known to be one of the magnetic axes and is hereafter called as $c$* \cite{kobets2009}.
	
	The ordering temperature of \MPS is found to be $ T_N $ = 77 K \cite{shemerliuk2021}. In contrast, the transition temperature of \MNPS is rather uncertain and might depend on the direction of the applied magnetic field. Various studies have reported different values of $T_N$, for instance it amounts to 12 K in \cite{Basnet2021}, 38 K in \cite{shemerliuk2021}, 41\,K in \cite{Yan2011} and 42 K in \cite{Lu2022}. For the samples used in this study the ordering temperatures were extracted from the temperature dependence of the susceptibility $\chi$ measured at $H = 1000$\,Oe (see Appendix, Fig.~\ref{fig:Mag_Ang_dep_MN} (a)). The  calculation of the maximum value of the derivative  $d (\chi \cdot T)/dT$ yields $T_N \sim$ 57 K for \HIIc and $T_N \sim$ 76 K for \HIIab (hereafter called $T_N$*).
	
	The antiferromagnetic resonance (AFMR) and ESR measurements (hereafter called HF-ESR) were performed on several single crystalline samples of \MPS and \MNPS\ using a homemade HF-ESR spectrometer. A superconducting magnet from Oxford instruments with a variable temperature insert (VTI) was used to generate magnetic fields up to 16 T allowing a continuous field sweep. The sample was mounted on a probe head which was then inserted into the VTI immersed in a $ ^{4} $He cryostat. A piezoelectric step-motor based sample holder was used for angular dependent measurements. Continuous He gas flow was utilized to attain stable temperatures in the range of 3 to 300 K. Generation and detection of microwaves was performed using a vector network analyzer (PNA-X) from Keysight Technologies. Equipped with the frequency extensions from Virginia Diodes, Inc., the PNA-X can generate a frequency in the range from 75 to 330 GHz. The measurements were performed in the transmission mode, where the microwaves were directed to the sample using oversized waveguides. All measurements were made by sweeping the field from 0 to 16 T and back to 0 T at constant temperature and frequency.
	
	\begin{figure}
		\centering
		\includegraphics[width=\linewidth]{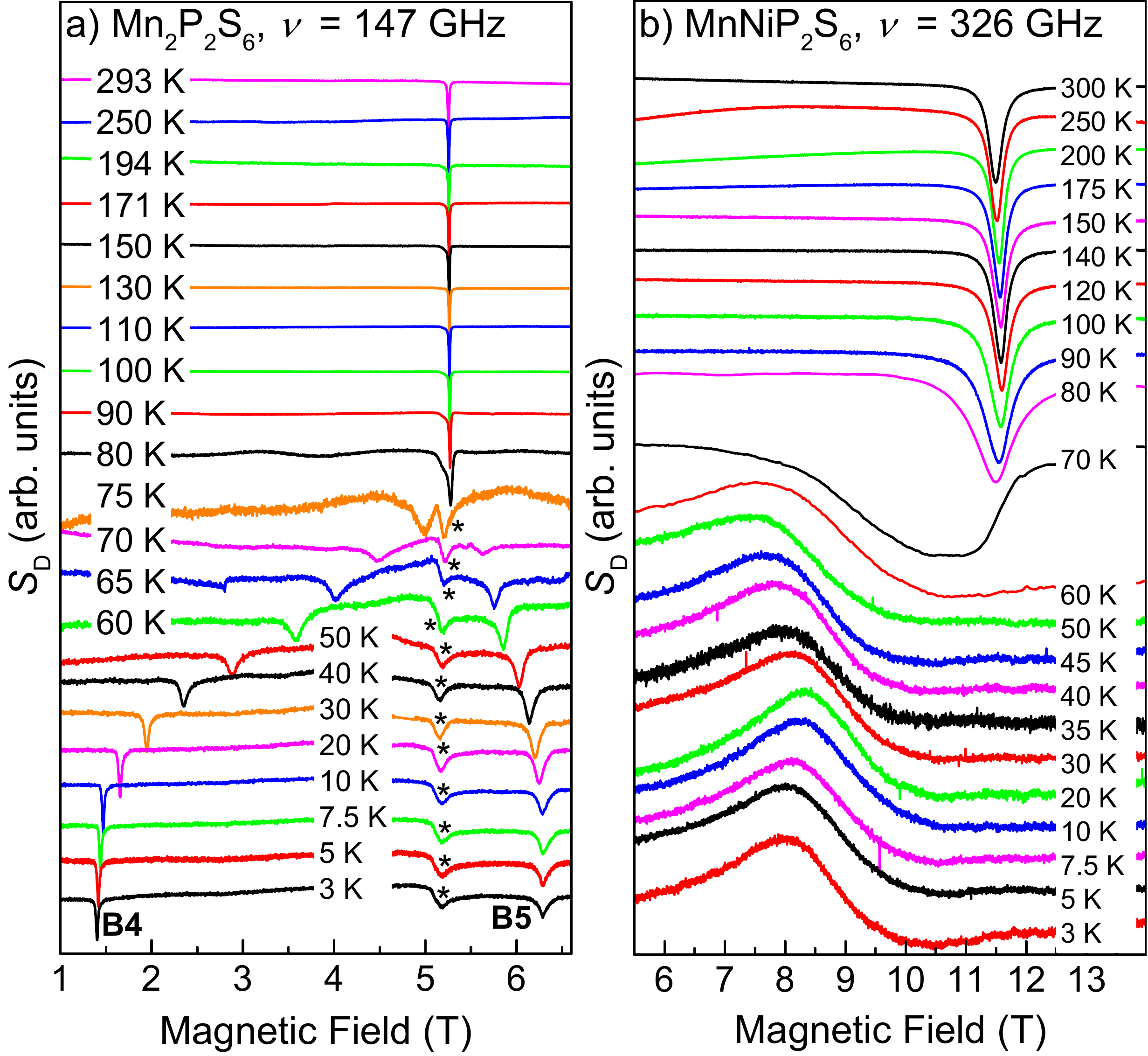}
		\caption{Temperature dependence of HF-ESR spectra of (a) \MPS at fixed excitation frequency $\nu \approx 147$ GHz and (b) \MNPS at $\nu \approx 326$ GHz in \HIIc configuration. Spectra are normalized and vertically shifted for clarity. The temperature independent peaks from the impurity in the probehead occurring at low frequencies are marked with asterisks.}
		\label{fig:T_Dep_H_para_c}
	\end{figure}
	
	HF-ESR signals generally have a Lorentzian line profile with an absorption and dispersion components. For such a case, the resonance field ($H_{res}$) and linewidth (full width at half maxima, $\Delta H$) can be extracted by fitting the signal with the function: 	
	
	\begin{align}
		S_D (H) = \ & \frac{2Amp}{\pi}\times(L_1sin\alpha+ L_2cos\alpha)\nonumber\\
		+ \ & C_{offset}+C_{slope}H 
		\label{eq:lorentz}
	\end{align}
	
	where $S_D (H)$ is the signal at the detector and $Amp$ is the amplitude. $C_{offset}$ represents the offset and $C_{slope}H$ is the linear background of the spectra. $L_1$ is the Lorentzian absorption which is defined in terms of $H_{res}$ and $\Delta H$. $L_2$ is the Lorentzian dispersion which is obtained by applying the Kramers-Kronig transformation to $L_1$. $\alpha$ is a parameter used to define the degree of instrumental mixing of the absorption and dispersion components which is unavoidable in the used setup. Some of the HF-ESR signals of \MPS could not be fitted using the above equation due to the development of the shoulders or the splitting of peaks \footnote{The splitting of the signal was mostly observed at high microwaves frequencies, at which the wavelength becomes comparable to the size of the measured sample flake. In such regime one could expect some instrumental effects distorting the line shape of the ESR signal.}. In this case $H_{res}$ and $\Delta H$ were obtained by picking the field position of the peak and by calculating the full width at half maximum of this peak, respectively.

\section{Results}
\label{sec:results}

	\subsection{Temperature dependence of HF-ESR response}

	To study the temperature evolution of the spin dynamics, the HF-ESR spectra were measured at several temperatures in the range of 3 - 300 K and at few selected microwave excitation frequencies $\nu$. Such dependences measured in the \HIIc configuration at $\nu = 147$\,GHz for \MPS and at $\nu = 326$\,GHz for \MNPS are presented in Fig.~\ref{fig:T_Dep_H_para_c}. As can be seen, in the case of \MPS upon entering the ordered state with lowering temperature, the single ESR line transforms into two modes B4 and B5 (see below) at $\nu = 147$\,GHz. The temperature dependence of the spectra for other frequencies can be found in Appendix in Fig.~\ref{fig:Tdep_other}.

	\begin{figure}[t]
		\centering
		\includegraphics[width=\linewidth]{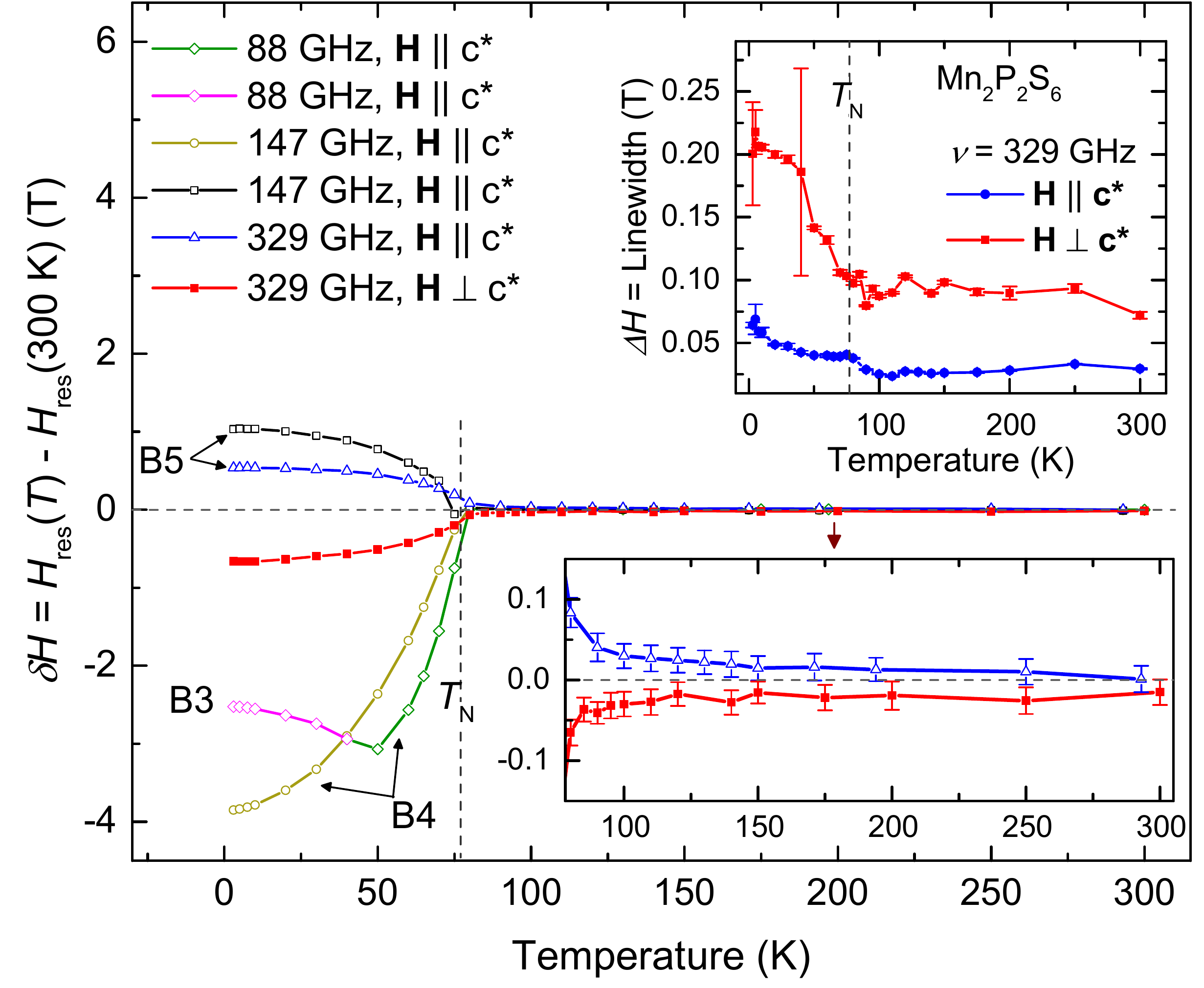}
		\caption{Shift of the resonance field position $\delta H$ (main panel) and linewidth $\Delta H$ (upper inset) as a function of temperature. The horizontal dashed line represents zero shift from the room temperature value and the vertical dashed line (also for inset) represents the Néel temperature of the material.}
		\label{fig:MPS_Hres}
	\end{figure}

	\begin{figure}[t]
	\centering
	\includegraphics[width=\linewidth]{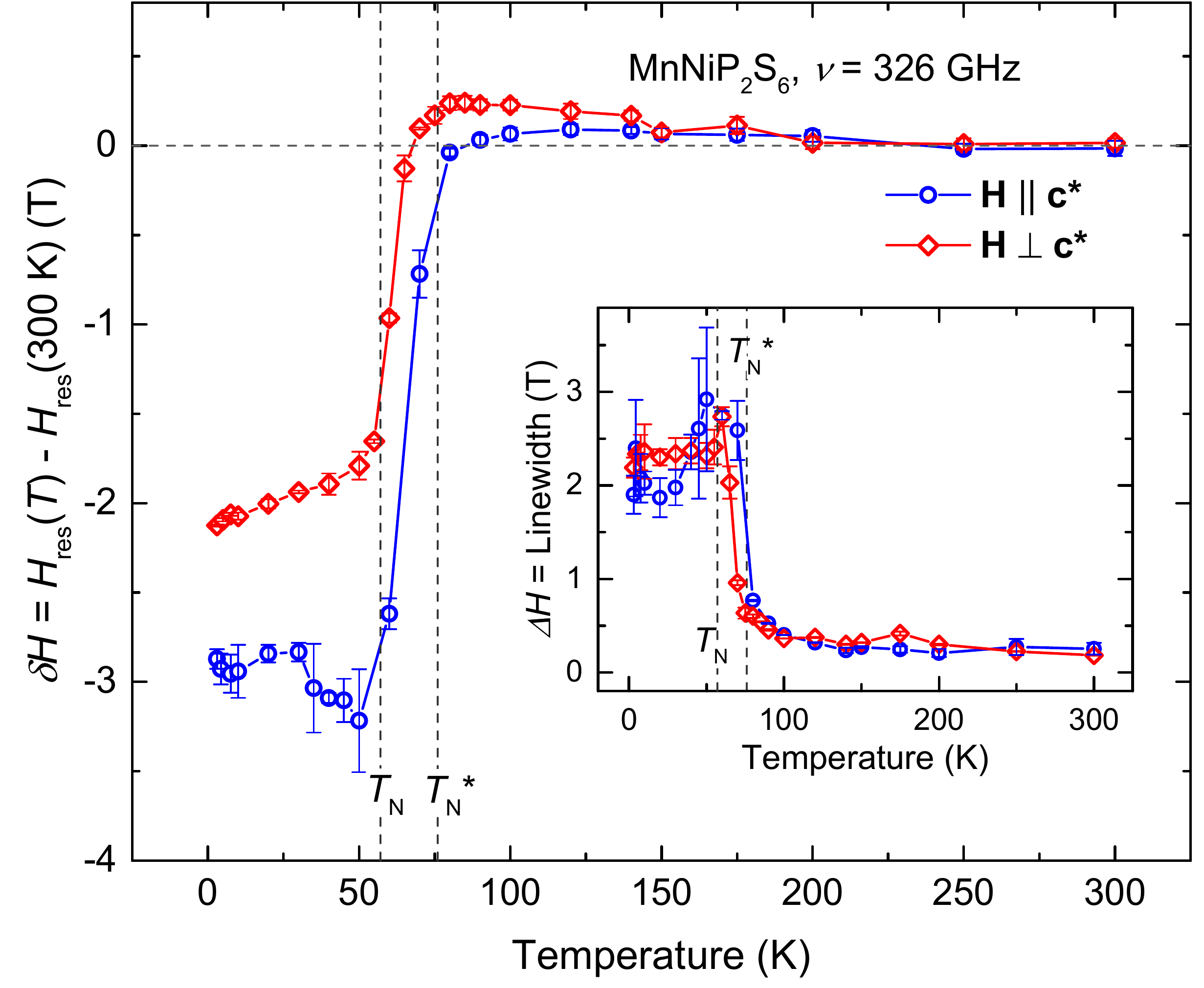}
	\caption{Temperature dependence of $ \delta H $ (main panel) and $ \Delta H $ (inset) measured at $\nu$ = 325.67 GHz. The horizontal dashed line represents zero shift from room temperature value and the vertical dashed line (also for inset) represents the Néel temperature of the material.}
	\label{fig:MNPS_Hres}
	\end{figure}
	
	The shift of the obtained values of $H_{res}$ from the resonance field position at $T = 300$\,K, $\delta H$ = $H_{res}$ - $H_{res}$(300 K) is plotted as a function of temperature for \MPS and \MNPS in Fig.~\ref{fig:MPS_Hres} and Fig.~\ref{fig:MNPS_Hres}, respectively. $H_{res}$(300 K) was calculated using the equation $h\nu = g\mu_{B}\mu_{0}H_{res}$, where $h$ is the Plank constant, $\mu_B$ is the Bohr magneton, $\mu_0$ is the permeability of free space and $g$ is the $g$-factor of resonating spins. The g-factor is obtained from the frequency dependence of the resonance field at 300 K (see Sec.~\ref{sec:Fdep_300K}). In the case of \MPS, $\delta H$ stays practically constant down to $T \sim 130 - 150 $\, K for both configurations \HIIc and \HIIab. Below this temperature it starts to slightly deviate (lower inset in Fig.~\ref{fig:MPS_Hres}), suggesting a development of the static on the ESR time scale internal fields. In contrast, the deviations of $\delta H$ from zero value in \MNPS are larger, and are observed at a higher temperature $T \sim 200$\,K. In the vicinity of the ordering temperature $T_N$* there is a strong shift of the ESR line, observed for both compounds. In the \MPS case the sign of $\delta H$ below the ordering temperature depends on the particular AFMR mode, which is probed at the specific frequency. This is detailed in the following Sec.~\ref{sec:Fdep_3K}. 
		
	Insets of Fig.~\ref{fig:MPS_Hres} and Fig.~\ref{fig:MNPS_Hres} represent the evolution of the linewidth $\Delta H$ as a function of temperature for \MPS and \MNPS compounds, respectively. At $T>T_N$, $\Delta H$ remains practically temperature independent for both compounds. A small broadening of the line is observed in the vicinity of the phase transition temperature, and there is a drastic increase of $ \Delta H$ in the ordered state. Note, that $\Delta H$ of \MNPS is larger than that of \MPS in the whole temperature range. Moreover, for \MNPS, $\Delta H$ increases at low temperatures by almost one order of magnitude from 0.3 to 3 T (inset in Fig.~\ref{fig:MNPS_Hres}). Such extensive line broadening at low temperatures hampers the accurate determination of the linewidth and resonance field, which is accounted for in the error bars.

	\subsection{Frequency dependence at 300 K}
	\label{sec:Fdep_300K}
	
	The frequency dependence of the resonance field $\nu(H_{res})$ of \MPS and \MNPS compounds measured in the paramagnetic state at $T$ = 300 K is shown in Fig.~\ref{fig:Fdep_300K}. Both plots have a linear dependence which can be fitted with the conventional paramagnetic resonance condition for a gapless excitation $h\nu = g\mu_{B}\mu_{0}H_{res}$. For \MPS, we obtain almost isotropic values of the $ g $-factor: $g_\parallel = 1.992 \pm 0.001$ (\textbf{H} $\parallel$ \textbf{c*}) and $g_\perp = 1.999 \pm 0.001$ (\textbf{H} $\perp$ \textbf{c*}), which is expected for a Mn$^{2+}$ ion \cite{abragam2012}. In contrast, \MNPS shows a small anisotropy of $ g $-factors with $g_{\parallel}$ = 2.026 $\pm$ 0.002 and $g_{\perp}$ = 2.047 $\pm$ 0.004. In case of Ni$ ^{2+} $ ions (3$d^{8}$, $S$ = 1), $g$-factors are expected to be appreciably greater than the free electron spin value, as is revealed in the HF-ESR studies on \NPS \cite{kavita2022}.

\subsection{Frequency dependence at 3 K}
\label{sec:Fdep_3K}
	
\subsubsection{\MPS}
\label{sec:Fdep_3K_MPS}

	\begin{figure}[t]
		\centering
		\includegraphics[width=\linewidth]{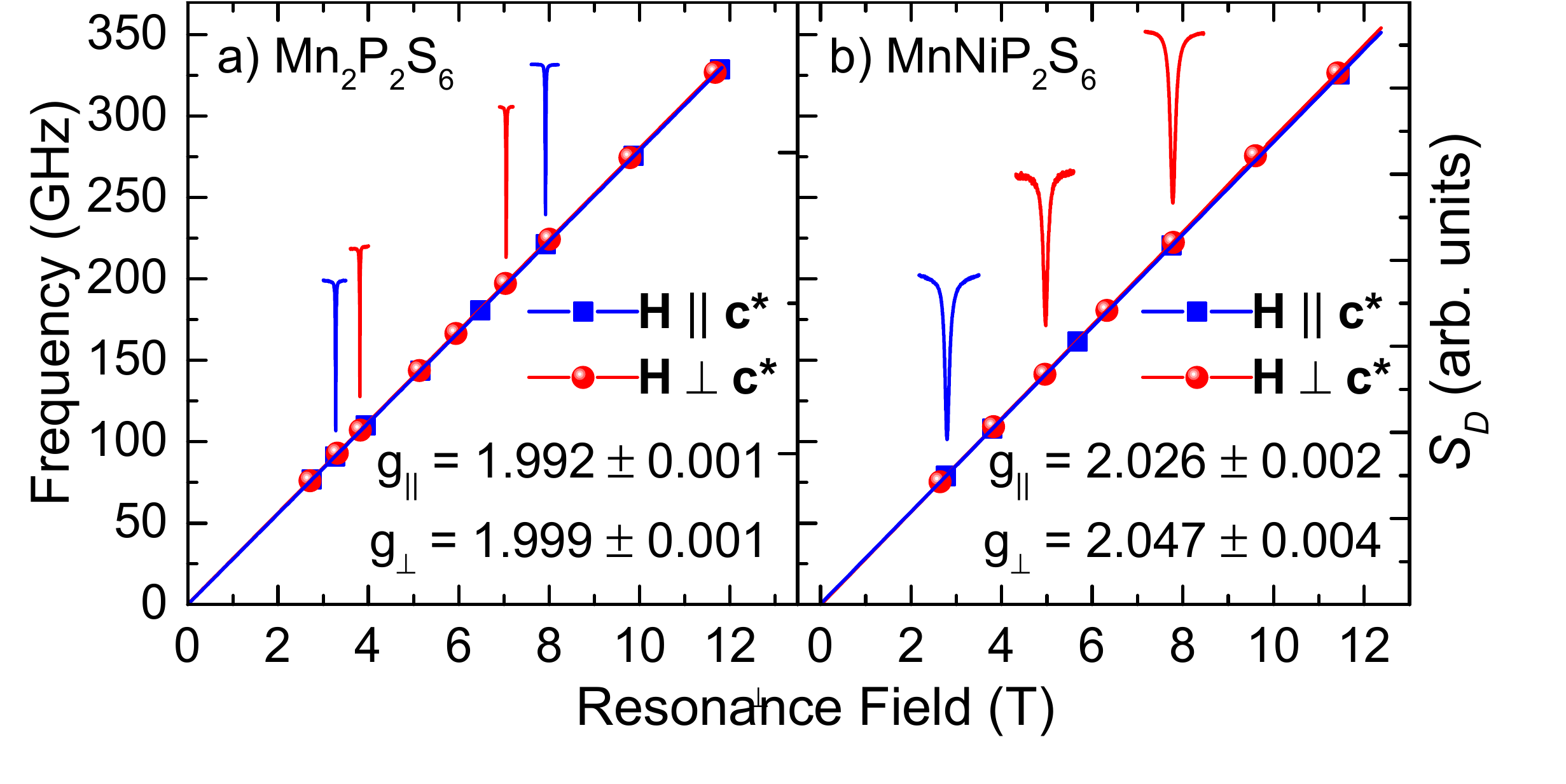}
		\caption{$\nu(H_{res})$ dependence measured at 300\,K for (a) \MPS and (b) \MNPS. Blue squares represent \HIIc configuration and the red circles represent \HIIab configuration. Solid lines show the results of the fit according to the resonance condition of a conventional paramagnet $h\nu = g\mu_{B}\mu_{0}H_{res}$. Right vertical axis: Representative spectra normalized for clarity. The color of the spectra corresponds to the color of the data points in the $\nu(H_{res})$ plot with the same $H_{res}$.}
		\label{fig:Fdep_300K}
	\end{figure}

	\begin{figure}[t]
		\centering
		\includegraphics[width=\linewidth]{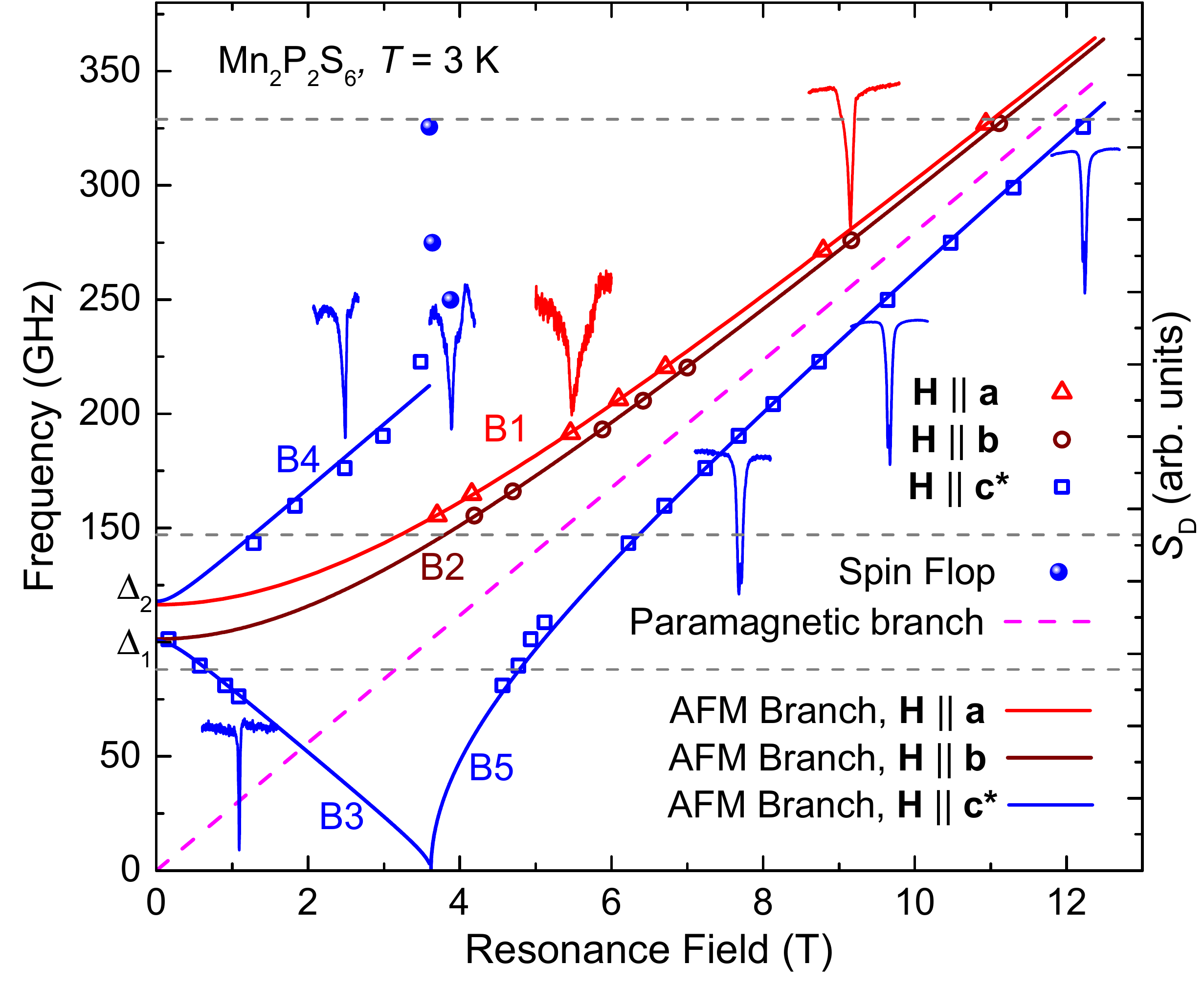}
		\caption{$\nu(H_{res})$ dependence of HF-ESR signals measured at $ \textit{T} $ = 3 K (symbols). Solid lines are the fit to the phenomenological equations as explained in the text. The dash gray lines correspond to the frequencies at which temperature dependent measurements were performed. The dash line in magenta represents the paramagnetic branch. Right vertical scale: Normalized ESR spectra for selected frequencies. For clarity the spectra are shifted vertically. Error bars in the $H_{res}$ are smaller than the symbol size.}
		\label{fig:nu(H)_3K_M}
	\end{figure}
		
\begin{figure}
			\centering
			\includegraphics[width=\linewidth]{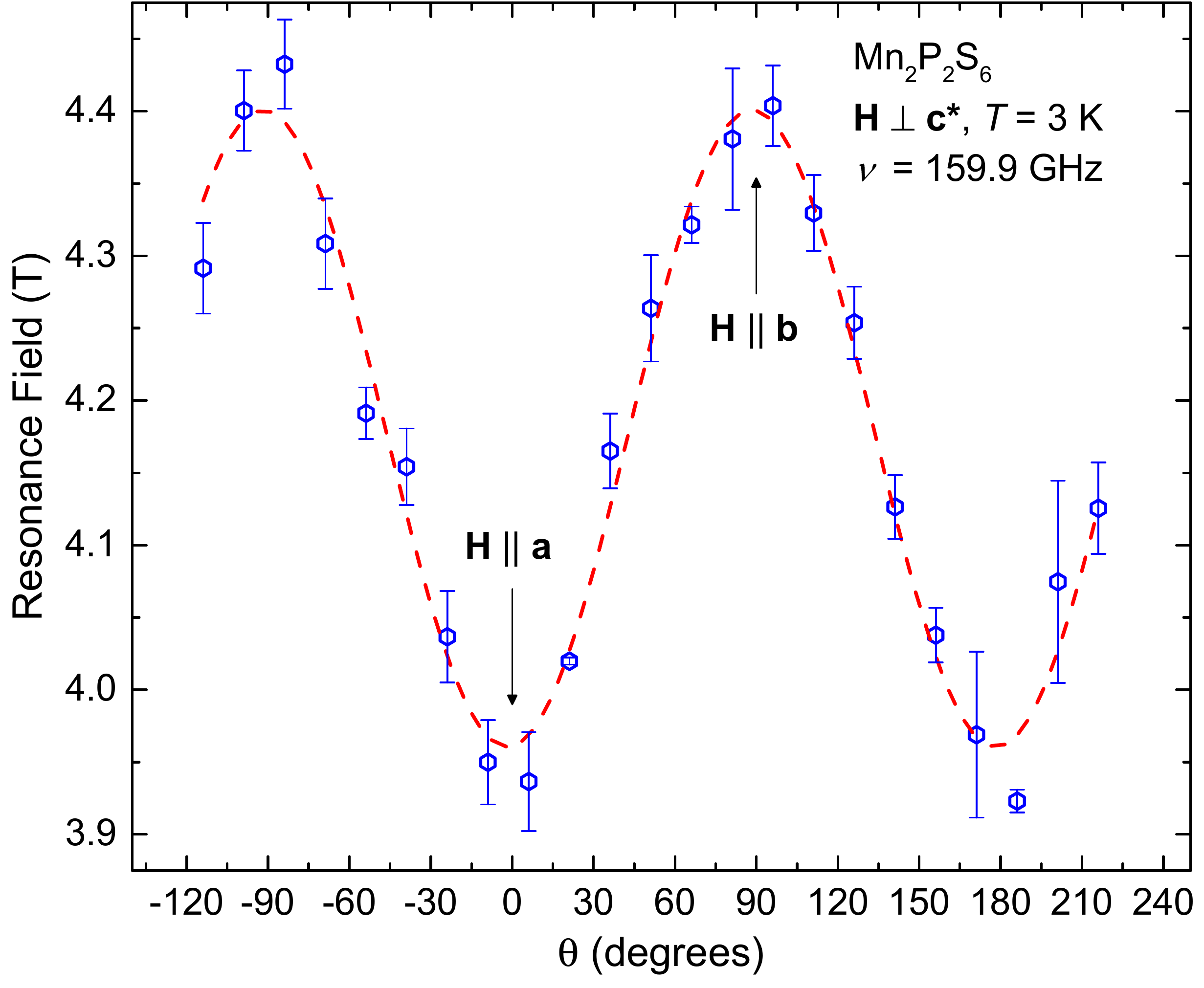}
			\caption{Resonance field as a function of angle $\theta$ at $T$ = 3 K and $\nu$ = 160 GHz for \MPS. $\theta$ denotes the angle between the direction of the field applied along the $ab$-plane and the $a$-axis. Red dash line represents the result of the fit, as explained in the text.}
			\label{fig:MPS_Ang_dep}
		\end{figure}

	The low temperature resonance modes of \MPS obtained at $T$ = 3 K are plotted in Fig.~\ref{fig:nu(H)_3K_M}. The measurements in the \HIIc configuration (Fig.~\ref{fig:nu(H)_3K_M}) yield three branches B3, B4 and B5, two of which (B3 and B4) are observed below the spin-flop field, ${H}_{SF}$ = 3.62 T. Branches B1 and B2 are assigned to the measurements along $ a $- and $ b $-axis, respectively \footnote{As it is suggested in Ref.~\cite{Goossens2010}, the more energetically favorable in-plane direction is $b$. In the case of an antiferromagnet, a given magnetic field applied along the $b$-axis yields a lower magnetization value, and, therefore, the internal field in such configuration is smaller compared to $\textbf{H} \parallel a$. This further implies that a larger external field is required to reach the resonance condition for $\textbf{H} \parallel b$ configuration. In the experiment, the magnetic field is applied at various angles by rotating the crystal in the \HIIab configuration. The orientation which yields the maximum $H_{res}$ is therefore $\textbf{H} \parallel b$. The frequency dependence was performed in that orientation to record mode B2. Similarly, branch B1 was recorded when $H_{res}$ had the minimum value (see Fig.~\ref{fig:MPS_Ang_dep}).}. Additionally, at the spin-flop field, a non-resonance absorption peak (full circles) was observed at high frequencies.

	The exact gap values were calculated by fitting the in-plane resonance branches B1 and B2 using the analytical expressions for easy-axis AFMs \cite{Turov}:

		\begin{equation}
			h\nu = [(g_{\perp} \mu_{B}\mu_{0}H_{res})^{2}+ \Delta_{1,2}^{2}]^{1/2}.
			\label{eq:B1}
		\end{equation}
		
	\noindent Here $ \Delta_{1}$ corresponds to the magnon excitation gap for branch B2 (also B3), and $\Delta_{2} $ corresponds to B1 (also B4). The obtained values are $ \Delta_{1} = \Delta_1^{\text{\MPS}} = 101.3 \pm 0.6$\,GHz and $\Delta_{2} = \Delta_2^{\text{\MPS}} = 116 \pm 2$\,GHz. These values, which agree well with previous measurements by \citet{Okuda1986} and \citet{kobets2009}, are then used in the theoretical description for a rhombic biaxial two-lattice AFM \cite{nagamiya1955,kobets2009} to match the field dependence of B3 and B4 \footnote{Considering the two magnetic sublattice AFM model, $\boldsymbol{M_1}$ and $\boldsymbol{M_2}$ are the sublattice magnetization vectors coupled antiparallely. When the field is applied along the easy-axis, these vectors start precessing around \textbf{H} clockwise or anticlockwise giving rise to two low field branches, B3 and B4. At the spin-flop field, the antiferromagnetic phase is no more stable and spins flop to the $b$ direction in the $ab$-plane, making an equal angle with the field. The mutual precession of flipped $\boldsymbol{M_1}$ and $\boldsymbol{M_2}$ vectors gives rise to the third branch B5 which increases with field \cite {rezende2019}}: 
		
		\begin{align}
			\nu = \frac{g\mu_{B}\mu_{0}}{\sqrt{2}h}\times& \Bigl(\Delta_{1}^{2} + \Delta_{2}^{2}+2H_{res}^{2} \pm \nonumber  \\
			\pm& \sqrt{8H_{res}^{2}(\Delta_{1}^{2} + \Delta_{2}^{2})+(\Delta_{1}^{2} - \Delta_{2}^{2})^{2}}\Bigl)^{1/2}.
			\label{eq:B4}
		\end{align}
		
		Above the spin-flop field the above model can not be used to describe the system. Therefore branch B5 \footnotemark[3] was simulated by the resonance condition of a conventional easy-axis AFM \cite{Turov}: 
		
		\begin{equation}
			h\nu = [(g_{\parallel} \mu_{B}\mu_{0}H_{res})^{2}- \Delta_{1}^{2}]^{1/2}.
			\label{eq:B5}
		\end{equation}
		
		The presence of the second easy-axis within the $ab$-plane is further confirmed by the angular dependence of $ H_{res}(\theta) $ in the \HIIab configuration (Fig.~\ref{fig:MPS_Ang_dep}). It follows a $ A + B sin^{2} (\theta)$ law, which suggests a 180$ ^{\circ} $ periodicity of $ H_{res}(\theta) $. $\theta$ denotes the angle between the applied field and $a$-axis. For a honeycomb spin system with a Néel type arrangement, a six-fold periodicity of angular dependence in the layer plane can be expected. However, this is absent in the case of \MPS sample due to dominating effects of a two-fold in-plane anisotropy.
		
		To further analyze the measured $\nu(H_{res})$ dependence of the AFMR modes in the magnetically ordered state of \MPS, that correspond to the collective excitations of the spin lattice (spin waves), we employed a linear spin wave theory (LSWT) with the second quantization formalism \cite{Turov,Holstein1940}. The details of our model are provided in Ref.~\cite{Alfonsov2021b}. The phenomenological Hamiltonian for the two-sublattice spin system, used for calculations of the spin waves energies, has the following form: 
		
		\begin{align}
			\label{Hamil}
			\mathscr{H} = \ & A \frac{(\boldsymbol{M_1 M_2})}{M_0^2} + K_{uniax} \frac{{M_{1_z}}^2 + {M_{2_z}}^2}{M_0^2} \nonumber \\
			+ \ & \frac{K_{biax}}{2} \frac{(M_{1_x}^2 - M_{1_y}^2) + (M_{2_x}^2 - M_{2_y}^2)}{M_0^2} \nonumber \\
			- \ & (\boldsymbol{\textbf{H} M_1}) - (\boldsymbol{\textbf{H} M_2})\, .
		\end{align} 
		
		Here the first term represents the exchange interaction between the magnetic sublattices with respective magnetizations $\boldsymbol{M_1 \ \text{and} \ M_2}$, such that $\boldsymbol{M_1}^2 = \boldsymbol{M_2}^2 = (M_0)^2 = (M_s/2)^2$, with $M_s^2$ being the square of the saturation magnetization. $A$ is the mean-field antiferromagnetic exchange constant. The second term in Eq.~(\ref{Hamil}) is the uniaxial part of the magnetocrystalline anisotropy given by the anisotropy constant $K_{uniax}$. The third term describes an additional anisotropy in the $xy$-plane with the respective constant $K_{biax}$. The fourth and fifth terms are the Zeeman interactions for both sublattice magnetizations. The results of the calculation match well with the measured data. In the calculation we assumed a full Mn saturation moment of $\sim5\mu_B$, yielding $M_s = 446$\,erg/(G$\cdot$cm$^3$) $= 446\cdot10^3$\,J/(T$\cdot$m$^3$), considering 4 Mn ions in the unit cell. The average $g$-factor value of 1.995 was taken from the frequency dependence measurements at $T = 300$\,K (Fig.~\ref{fig:Fdep_300K}). As the result we obtain the exchange constant $A= 2.53\cdot10^8$\,erg/cm$^3$ $= 2.53\cdot10^7$\,J/m$^3$, uniaxial anisotropy constant $K_{uniax} = -7.2\cdot10^4$\,erg/cm$^3$ $= -7.2\cdot10^3$\,J/m$^3$, and an in-plane anisotropy constant $K_{biax} = 1.9\cdot10^4$\,erg/cm$^3$ $= 1.9\cdot10^3$\,J/m$^3$. Within the mean-field theory $A$ is related to the Weiss constant $\Theta = A*C/M_0^2$, where $C$ is the Curie constant. $\Theta$, that provides an average energy scale for the exchange interaction in the system, amounts therefore at least to $\Theta_{\text{\MPS}} \approx 350$\,K.

	\subsubsection{\MNPS}

\begin{figure}
	\centering
	\includegraphics[width=\linewidth]{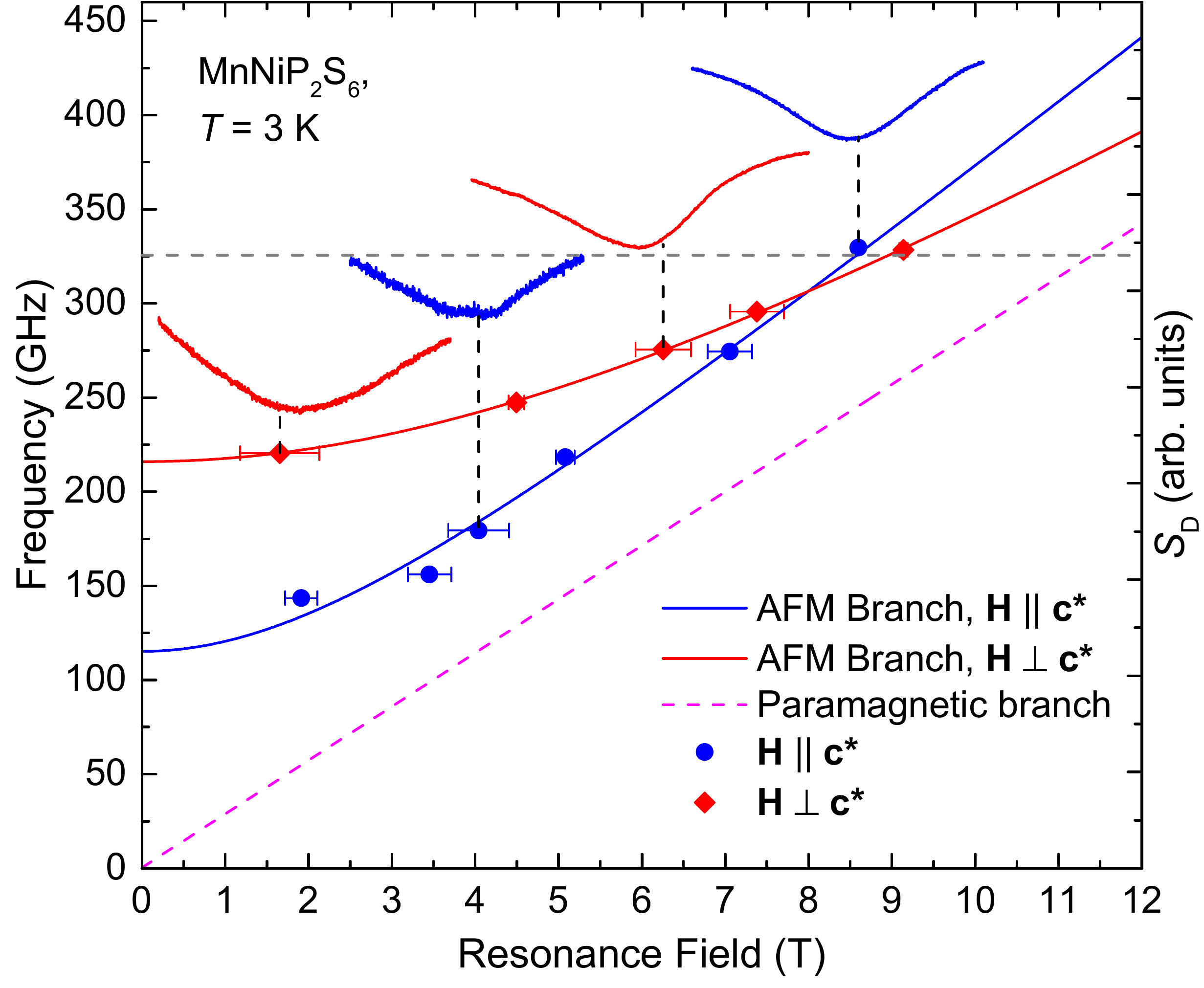}
	\caption{$\nu(H_{res})$ dependence for \MNPS measured at 3 K for both configurations of magnetic field. Right vertical scale: Exemplary spectra positioned above the resonance points. The horizontal dash gray line represents the frequency at which the temperature dependence was measured. The dash line in magenta depicts the paramagnetic resonance branch at 300 K.}
	\label{fig:nu(H)_3K_MN}
\end{figure}
	
	In the case of \MNPS we observe one branch for \HIIc and another one for \HIIab configuration, respectively, as shown in Fig.~\ref{fig:nu(H)_3K_MN}. HF-ESR spectra were also recorded at various angles for the in-plane orientation. Within the experimental error bars of $\sim 300$\,mT, no signatures for an in-plane anisotropy were observed (see Fig.~\ref{fig:Mag_Ang_dep_MN} in Appendix). Both branches follow the resonance condition for a hard direction of an AFM given by Eq.~(\ref{eq:B1}), which reveals that neither $c$*-axis nor the $ab$-plane are energetically favorable. The magnitude of the gap was obtained from the fit as $\Delta_1^{\text{\MNPS}} = 115 \pm 9 $\,GHz for \HIIc and $\Delta_2^{\text{\MNPS}} = 215 \pm 1 $\,GHz for \HIIab configurations, respectively.
	
	Unfortunately, we could not find a good matching of the calculated frequency dependence to the one measured at low temperature (Fig.~\ref{fig:nu(H)_3K_MN}) with the AFM Hamiltonian for a two sublattice model. Inclusion of the terms describing cubic, hexagonal and symmetric exchange anisotropies in addition to those given in Eq.~(\ref{Hamil}) did not yield a good result either. This could be explained by the complicated type of order of two magnetically inequivalent ions Mn$^{2+}$ ($S = \frac{5}{2}$, $g = 1.955$) and Ni$^{2+}$ ($S = 1$, $g = 2.17$), which possibly requires a more sophisticated model than the one used in this study. The analysis might be even more complicated by potential disorder in the system due to the stochastic distribution of these ions on the 4g Wyckoff sites. Therefore the full description of this system remains an open question. However, one could draw some conclusions by analyzing how the magnetization measured at low T depends on the Mn/Ni ratio \cite{shemerliuk2021}. The reduction of the magnetization measured at low T can be explained by the reduction of the total moment per formula unit of \MNPS, which can be found as an average of the Mn and Ni saturation magnetizations and amounts to $\sim 7.2\,\mu_B$, compared to \MPS which has the saturation moment of $\sim 10\,\mu_B$. Additionally, an almost isotropic behavior of the magnetization as a function of magnetic field (inset of Fig.~\ref{fig:Mag_Ang_dep_MN} (a)) suggests that the isotropic exchange energy is by orders of magnitude the strongest term defining the static magnetic properties of \MNPS. In this case, the magnetization value, measured at the magnetic field applied along some hard direction, should be inversely proportional to the mean-field isotropic exchange constant $M \sim H/A$. The reduced magnetization in \MNPS suggests, therefore, that $\Theta \sim A$ should be at least as large in \MNPS ($\Theta_{\text{\MNPS}} \ge \text{\,} \sim 350$\, K) as in \MPS ($\Theta_{\text{\MPS}} \approx 350$\,K, see Sec.~\ref{sec:Fdep_3K_MPS}).

\section{Discussion}

	\subsection{Spin-Spin correlations in \PS ($T > T_N\text{*}$)}

		As has been shown in our previous work, both the resonance field and the linewidth of the HF-ESR signal in \NPS remain temperature independent by cooling the sample down to temperatures close to $T_N$ \cite{kavita2022}. Usually, in the quasi-2D spin systems the ESR line broadening and shift occur at $T > T_N$ due to the growth of the in-plane spin-spin correlations resulting in a development of slowly fluctuating short-range order \cite{Benner1990}. Specifically, the slowly fluctuating spins produce a static on the ESR timescale field causing a shift of the resonance line, and a distribution of these local fields and shortening of the spin-spin relaxation time due to the slowing down of the spin fluctuations increase the ESR linewidth. In the \MPS compound these features are not very pronounced, only in the resonance field of the HF-ESR response one can detect within error bars small deviations starting at $T \sim 130 - 150 $\, K. In the \MNPS compound, in turn, the critical broadening and the shift of the resonance line are observed at temperature $T \sim 200$\,K, which is much higher than $T_N$. Even though the critical broadening and the line shift above $T_N$ are much stronger pronounced in \MNPS, our previous low-frequency ESR study shows that the clear cut signatures of 2D correlated spin dynamics are present above $T_N$ only in the \MPS compound \cite{senyk2022}. Interestingly, these signatures, seen in the characteristic angular dependence of the ESR linewidth, develop only at elevated temperatures, where the effect of the strong isotropic AFM coupling ($\Theta_{\text{\MPS}} \approx 350$\,K) on the spin fluctuations becomes gradually suppressed. Critical broadening and the shift of the ESR line in \MNPS above $T_N$ could therefore be due to the stochastic distribution of Mn and Ni ions on the 4g Wyckoff sites of the crystal structure causing a competition of different order types with contrasting magnetic anisotropies. Our conclusion on the drastic difference in the ground states is supported by the strong distinction in the energy gaps and magnetic field dependences of the low-T spin wave excitations in \MPS, \MNPS and \NPS, respectively. The competing types of magnetic order -- the out-of-plane N\'eel type in \MPS and the in-plane zig-zag type in \NPS\ -- meeting in \MNPS might enhance spin fluctuations seen in the HF-ESR response at elevated temperatures. Strong fluctuations suppress, in turn, the ordering temperature for \MNPS which is evident in the recent studies on the \PS series \cite{shemerliuk2021, Basnet2021, Lu2022}. Moreover, in this scenario of the stochastic distribution of Mn and Ni, small deviation of the stoichiometry from sample to sample of the same nominal composition could vary the ordering temperature, which explains the broad range of $T_N$ measured in \MNPS samples \cite{shemerliuk2021, Basnet2021, Lu2022}.

	\subsection{Ground state and anisotropy of \PS ($T << T_N\text{*}$)}
	\label{sec:discussion2}
	
		At the lowest measurement temperature \MPS has an antiferromagnetic ground state with biaxial type of anisotropy, and the spin wave excitations can be successfully modeled using LSWT. As the result we obtain the estimation of the exchange interaction $\Theta_{\text{\MPS}} \approx 350$\,K and the parameters of the anisotropy $K_{uniax} = -7.2\cdot10^4$\,erg/cm$^3$ $= -7.2\cdot10^3$\,J/m$^3$ and $K_{biax} = 1.9\cdot10^4$\,erg/cm$^3$ $= 1.9\cdot10^3$\,J/m$^3$. There is only about four times difference between $K_{uniax}$ and $K_{biax}$, which suggests that the anisotropy in the $ab$-plane makes a significant contribution to the properties of the ground state of \MPS. Interestingly, the value of $K_{biax} = 1.9\cdot10^3$\,J/m$^3$ $\approx 2\cdot10^{-25}$\,J/spin is very close to the estimation of the anisotropy within the $ab$-plane made by \citet{Goossens2010}, suggesting a possible dipolar nature of this anisotropy. In the \MNPS case we could not find an appropriate Hamiltonian within a two sublattice model which would fully describe the system, calling for a more sophisticated theoretical study. Interestingly, the characteristic feature of the \MNPS compound is the almost isotropic dependence of the magnetization as a function of magnetic field, measured at temperature well below $T_N$ \cite{shemerliuk2021}. The isothermal magnetization measurements made on the sample used in this study, confirm the presence of this almost isotropic static magnetic response (see appendix Fig.~\ref{fig:Mag_Ang_dep_MN} (a)). Such an isotropic behavior of the static magnetization is related to the strong isotropic AFM exchange interaction ($\Theta_{\text{\MNPS}} \ge \text{\,} \sim 350$\, K), which is larger than the Zeeman energy of the applied magnetic field and the observed magnetic anisotropy in this system. However, the HF-ESR data reveals a substantial anisotropy in the magnetic field dependence of the spin waves. This seeming contradiction is actually not surprising. The magnetization value at the magnetic field applied along some hard direction is mostly given by the mean-field exchange constant $M \sim H/A$, whereas the magnon gap measured in the ESR experiment is roughly proportional to the square root of the product of exchange and magnetic anisotropy constants \cite{Turov}. 
		
		Qualitatively, the evolution of the type of magnetic anisotropy with $x$ in \PS is also evident from our study, where, e.g., \MNPS reveals no easy-axis within or normal to the $ab$-plane. 
		In order to quantify the change of magnetic anisotropic properties with the Mn/Ni content the excitation energy gaps can be used. The single gap of about 260 GHz was found in our previous study on \NPS \cite{kavita2022}. Both Mn containing compounds have two gaps $\Delta_1^{\text{\MNPS}} = 115 \pm 9 $\,GHz and $\Delta_2^{\text{\MNPS}} = 215 \pm 1 $\,GHz in the case of \MNPS, and $\Delta_1^{\text{\MPS}} = 101.3 \pm 0.6 $\,GHz and $\Delta_2^{\text{\MPS}} = 116 \pm 2 $\,GHz in the case of \MPS. As can be seen, there is a noticeable increase of the zero field AFM gaps in the samples with higher Ni content, suggesting an increase of the magnetic anisotropy and exchange interaction. Indeed, the estimated energy scale of the exchange interaction in \MPS is about $\sim 350$ K, in \MNPS is more than $\sim 350$ K, and it is even larger in \NPS, due to the observation of the larger $T_N$. This is also suggested by the previous investigations \cite{senyk2022,Wildes2022,Lancon18,Kim2019}. Mn$^{2+}$ with the half filled 3d electronic shell, and a small admixture of the excited state $^4 P _{5/2}$ into the ground state $^6 S _{5/2}$ is an ion with rather isotropic magnetic properties. In contrast, the ground state of the Ni$^{2+}$ ion in the octahedral environment \cite{Wang2018a} is a spin triplet with the higher lying orbital multiplets, admixed through the spin-orbit coupling \cite{abragam2012}, which makes the Ni spin ($S = 1$) sensitive to the local crystal field. This, first, could increase a contribution of the local (single ion) magnetic anisotropy term in the Hamiltonian describing the system in the ordered and in the paramagnetic state, as discussed for the case of \NPS in \cite{kavita2022}. Second, it could yield a deviation of the g-factor from the free electron value and also induce an effective $g$-factor anisotropy. The effective $g$-factor value and its anisotropy, as found in our study, increase with Ni content. Deviation of the $g$-factor from the free electron value ($\Delta g$) and the anisotropy of the exchange originate from the spin-orbit coupling effect, and therefore are interrelated. In the case of symmetric anisotropic exchange the elements of the anisotropic exchange tensor are $\mathcal{A} \propto (\Delta g / g)^2 J $ \cite{Kubo1954,Moriya1960,Kataev2001}, where $J$ is the isotropic exchange interaction constant. Observation of increased $\Delta g$ at higher Ni content suggests that in the Ni containing \PS the exchange anisotropy is likely an important contributor to the anisotropic properties of the ground state at low temperatures $< T_N$, such as, e.g., increased magnon gaps.

		\subsection{Critical behavior of \MPS ($T \lesssim T_N\text{*}$)}

		In the following we discuss the temperature dependence of the excitation energy gap $\Delta$ at finite magnetic fields in the collinear and the spin-flop AFM ordered phases of \MPS, at $H < H_{\rm sf}$ and $H > H_{\rm sf}$, respectively.  This should provide useful insights onto the type of the critical behavior of the  Mn spin lattice at $T < T_{\rm N}$. Such a dependence can be obtained by analyzing the temperature dependence of the shift of the resonance field positions $H_{\rm res}(T)$ of the excitation modes B4 and B5 for \HIIc\ (Fig.~\ref{fig:MPS_Hres}) with the aid of the simplified  relations $\Delta \approx h\nu - g_{\parallel} \mu_{B}\mu_{0}H_{res}$ for mode B4 and $\Delta \approx  [(g_{\parallel} \mu_{B}\mu_{0}H_{res})^2 - (h\nu)^2]^{1/2}$ for mode B5 derived from Eqs.~(\ref{eq:B4}) and (\ref{eq:B5}), respectively. 
		
		\begin{figure}
			\centering
			\includegraphics[width=\linewidth]{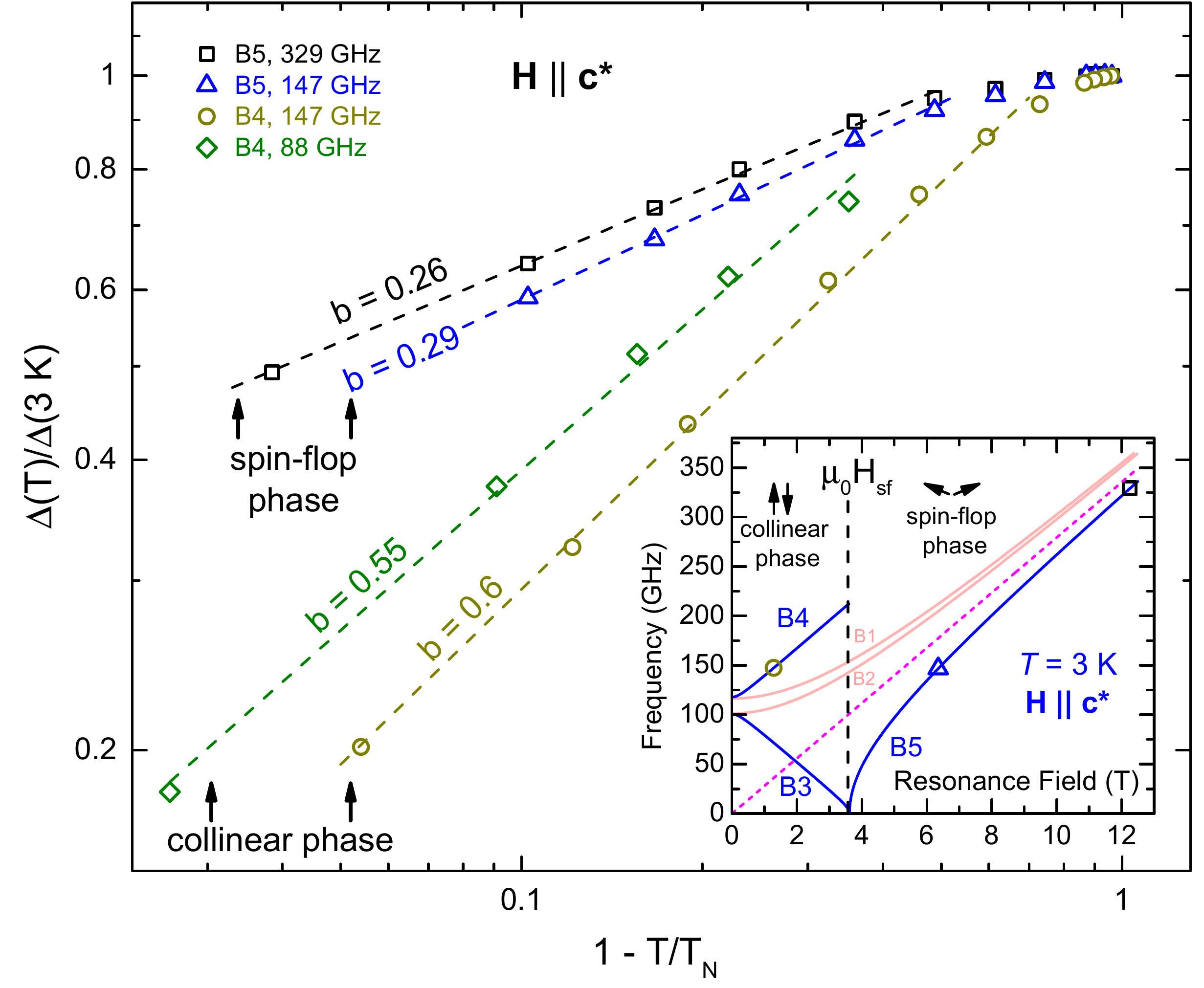}
			\caption{Main panel: Temperature dependence of the normalized energy gap $\Delta(T)/\Delta(3\,{\rm K}) = [1-(T/T_{\rm N})]^b$ for \MPS\ at different field regimes. Symbol shapes and colors correspond to that in Fig.~\ref{fig:MPS_Hres}. Inset: Resonance branches at $T = 3$\,K (solid lines) as in Fig.~\ref{fig:nu(H)_3K_M}. Symbols (same as in the main panel) indicate the positions of the resonance modes B4 at 147\,GHz and B5 at 147 and 329\,GHz.  The position of mode B4 at 88\,GHz is not shown here since it can be detected at $T \geq 50$\,K only. The temperature dependence of these modes shown in Fig.~\ref{fig:MPS_Hres} was used to estimate that of $\Delta(T)$. (see the text)	}
			\label{fig:exponents}
		\end{figure}
		
		The result of this analysis is shown in Fig.~\ref{fig:exponents}. The $\Delta(T)$ dependence can be well fitted to the power law $\Delta(T) \propto [1-(T/T_{\rm N})]^b$ in a broad temperature range below $T_{\rm N}$ with some deviations from it at lower $T$. The  exponents $b$ indicated in this Figure appear to be very different for modes B4 and B5. Notably, the resonance field of mode B4 is always smaller than the spin-flop field,  $H^{\rm B4}_{\rm res}\mid_ {\rm 88\,GHz} <  H^{\rm B4}_{\rm res}\mid_ {\rm 147\,GHz} < H_{\rm sf}$, whereas mode B5 occurs at larger fields with   $H_{\rm sf} < H^{\rm B5}_{\rm res}\mid_ {\rm 145\,GHz} <  H^{\rm B5}_{\rm res}\mid_ {\rm 329\,GHz}$ [Fig.~\ref{fig:exponents}(inset)]. This suggests a significant difference in the temperature dependence of the excitation gap in the collinear and spin-flop AFM ordered phases of \MPS.
		
		Usually, the magnetic anisotropy gap $\Delta(T)$ observed in quasi-2D antiferromagnets scales with the sublattice magnetization $M_{\rm sl}(T)$ \cite{Nagata1974,Uijen1984,Arts1990} so that the exponent $b$ of the temperature dependence  of $\Delta$ can be treated as a critical exponent $\beta$ of the AFM order parameter $M_{\rm sl}$.  If that were the case for \MPS, the value of $b$ in the collinear phase would indicate the mean-field behavior of $M_{\rm sl}(T)$ for which $\beta = 0.5$ (Fig.~\ref{fig:exponents}). In contrast, a strong reduction of $b$ in the spin-flop phase, as seen in Fig.~\ref{fig:exponents}, would correspond to the critical behavior of $M_{\rm sl}(T)$  in the 2D XY model for which $\beta = 0.231$ \cite{Bramwell93}.
		However, measurements of the temperature dependence of $M_{\rm sl}$ by elastic and of $\Delta$ by inelastic neutron scattering in zero magnetic field reveal a more complex scaling between these two parameters with $b \approx 3\beta/2$ and $\beta =0.32$ in the vicinity of $T_{\rm N}$, and $b \approx \beta$ with $\beta = 0.25$ at lower temperatures \cite{Wildes1998,Wildes2006,Wildes2007}. 
		
		This finding was tentatively ascribed to different temperature dependence of the competing single-ion and dipolar anisotropies which are both responsible for a finite value of $\Delta$ in the AFM ordered state of \MPS \cite{Wildes1998}. 
		Theoretical analysis in Ref.~\cite{Goossens2010} shows that besides the dipolar anisotropy which is responsible for the out-of-plane order of the Mn spins there is a competing, presumably single ion anisotropy turning the spins into the $ab$~plane. As argued in Ref.~\cite{Wildes2006}, the presence of the latter contribution  gives rise to the 2D XY critical behavior.

		It should also be noted that the scaling $b \approx 3\beta/2$ is a characteristics of a 3D antiferromagnet, as it follows from the theories of AFM resonance \cite{Nagamiya1951,Keffer1952,Kanamori1962} and was confirmed experimentally (see, e.g., \cite{Johnson1959,Richards1965}). Thus, a field-dependent change of $b$ indicates a kind of  field-driven dimensional crossover of the spin wave excitations at intermediate temperatures below $T_{\rm N}$ while ramping the magnetic field across the spin-flop transition.  Magnetic fields $H > H_{\rm sf}$ push the spins into the plane, boosting the effective XY anisotropy, which changes the character of spin wave excitations observed by ESR towards the 2D XY scaling regime. Interestingly, even if a strong field is applied in the crystal plane, then according to a recent nuclear magnetic resonance study of \MPS performed in a field of 7\,T \cite{Bougamha2022}, the exponent $\beta$ of the $T$ dependence of the internal field proportional to $M_{\rm sl}$ amounts to $\beta = 0.22$, close to the 2D XY value of 0.231.

\section{Conclusion}

In summary, we have performed a detailed ESR spectroscopic study of the single-crystalline samples of the van der Waals compounds \MPS and \MNPS. The measurements were carried out in a broad range of excitation frequencies and temperatures, and at different orientations of the magnetic field with respect to the sample. Our study suggests a strong sensitivity of the type of magnetic order and anisotropy below $T_N$, as well as of the g-factor and its anisotropy above $T_N$ to the Ni concentration. Stronger deviation of the g-factor from the free electron value in the samples containing Ni suggests that the anisotropy of the exchange can be an important contributor to the stabilization of the certain type of magnetic order with particular anisotropy. Analysis of the spin excitations at $T<<T_N$ has shown that both \MPS and \MNPS are strongly anisotropic. In fact, increasing the Ni content yields a larger magnon gap in the ordered state ($T<<T_N$). In the \MPS compound we could fully describe the magnetic excitations using a two sublattice AFM Hamiltonian, which yielded an estimation of the uniaxial anisotropy energy, the anisotropy energy within the $ab$-plane, and the average exchange interaction $\Theta_{\text{\MPS}} \approx 350$\,K. On the contrary, in the \MNPS compound the ground state and the excitations appear too complex to be described using two-sublattice AFM model. This could be due to a stochastic mixing of two magnetically inequivalent ions, Mn and Ni, on the 4g Wyckoff crystallographic sites. Nevertheless, the analysis of the magnetization measured at low-T suggests that the exchange coupling in this compound should be comparable to or stronger than that in \MPS. 

We have analyzed the spin-spin correlations resulting in a development of slowly fluctuating short-range order, which, in the quasi-2D spin systems, manifest in the ESR line broadening and shift at $T > T_N$. The line broadening and shift are much stronger pronounced in \MNPS compared to \MPS, suggesting that the critical broadening and the shift of the ESR line in \MNPS could be due to the enhanced spin fluctuations at the elevated temperatures caused by the competition of different types of magnetic order -- the out-of-plane N\'eel type in \MPS and the in-plane zig-zag type in \NPS. Moreover, these strong spin fluctuations in the mixed Mn/Ni compounds could additionally lower the ordering temperature.

Finally, the analysis of the temperature dependence of the spin excitation gap in \MPS at different applied fields suggests a kind of field-driven dimensional crossover of the spin wave excitations at intermediate temperatures below $T_{\rm N}$. Strong magnetic fields push the spins into the plane, boosting the effective XY anisotropy, which changes the character of spin wave excitations observed by ESR from a 3D-like towards the 2D XY scaling regime.

\begin{acknowledgments}
	
	J.J.A. acknowledges the valuable discussions with Kranthi Kumar Bestha. This work was supported by the Deutsche Forschungsgemeinschaft (DFG) through grants No. KA 1694/12-1, AL 1771/8-1, AS 523/4-1, and within the Collaborative Research Center SFB 1143 ``Correlated Magnetism – From Frustration to Topology'' (project-id 247310070), and the Dresden-Würzburg Cluster of Excellence (EXC 2147) ``ct.qmat - Complexity and Topology in Quantum Matter'' (project-id 390858490), as well as by the UKRATOP-project (funded by BMBF with Grant No. 01DK18002).

\end{acknowledgments}

\FloatBarrier

\onecolumngrid

\appendix*
\section{}

\begin{figure}[!h]
	\centering
	\includegraphics[width=0.49\columnwidth]{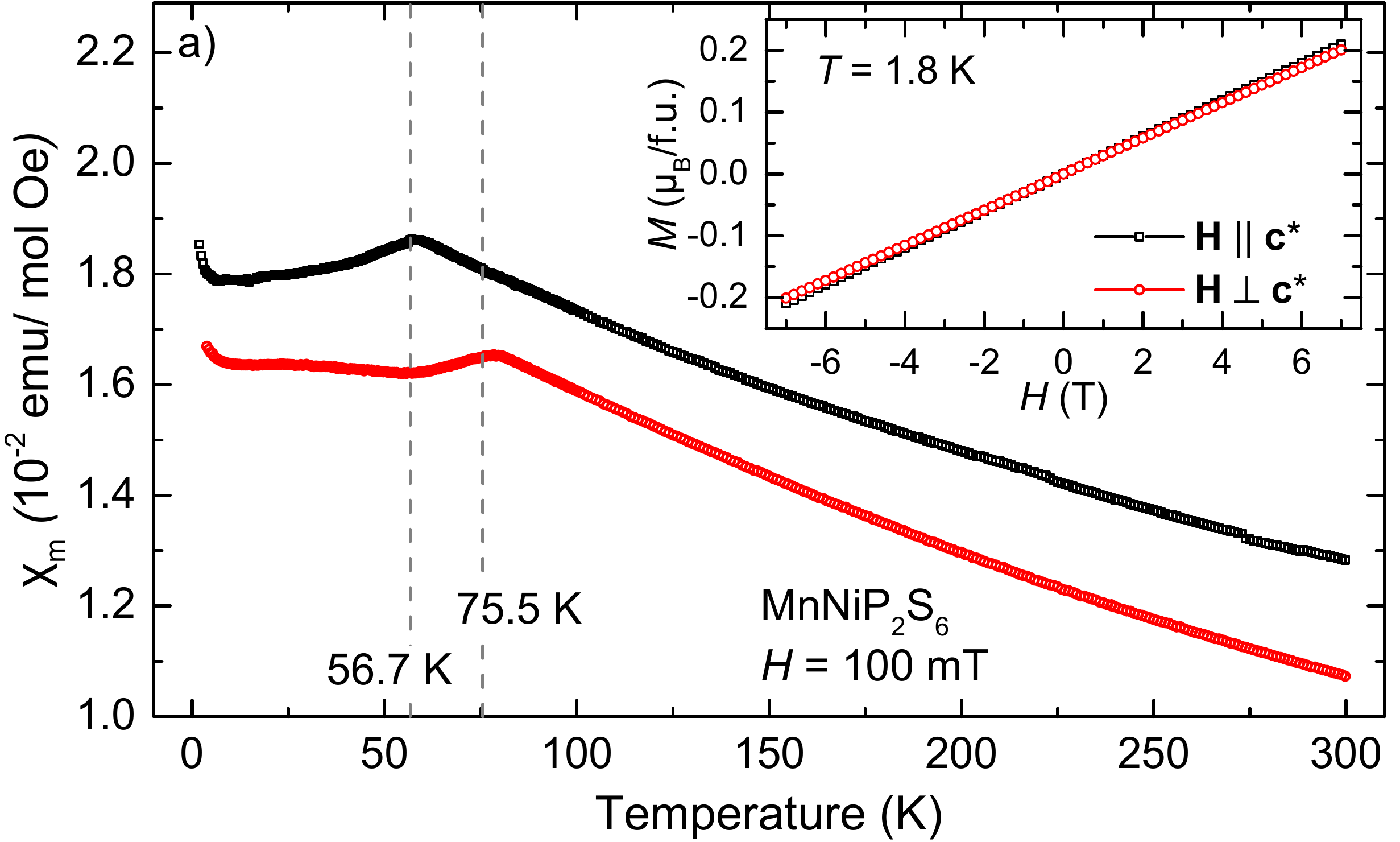}
	\includegraphics[width=0.49\columnwidth]{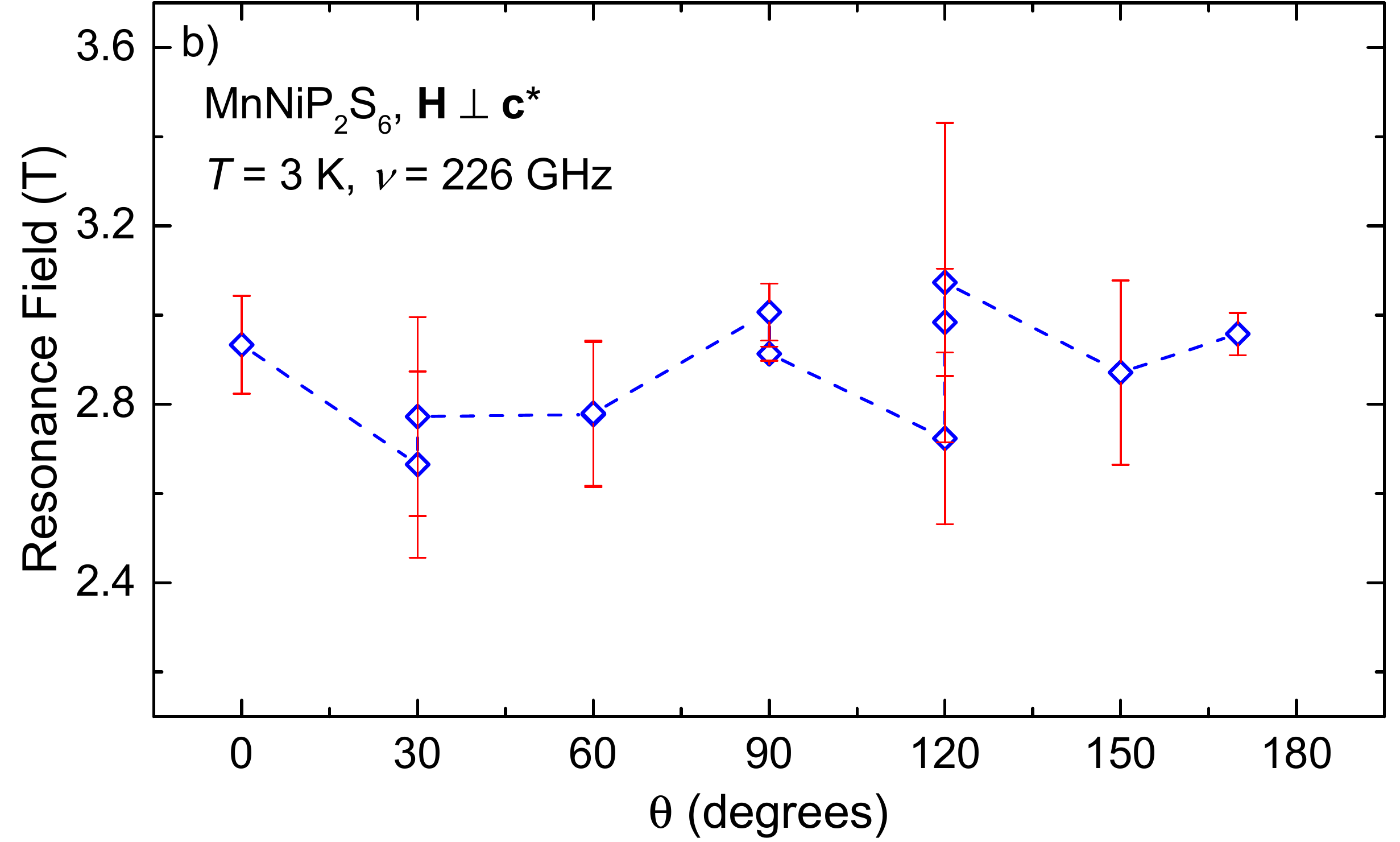}
	\caption{(a) Molar susceptibility at the applied field of 1000 Oe as a function of temperature measured on the sample of \MNPS, which was used for the ESR investigations. The gray dash lines represent the magnetic phase transition temperature in both configurations of the applied magnetic field. Inset: Isothermal magnetization per formula unit as a function of applied field measured at 1.8 K for \MNPS, depicting the almost isotropic field dependence of the magnetic response. (b) In-plane angular dependence of the resonance field at $T$ = 3 K and $\nu$ = 226 GHz for \MNPS, showing no systematic angular dependence within the average error bar of $\sim 0.16$\,T. The large linewidth values $\Delta H$ of the peaks in the ESR spectra are accounted for in the enlarged error bars.}
	\label{fig:Mag_Ang_dep_MN}
\end{figure}

\vspace{-10pt}

\begin{figure}[!h]
	\centering
	\includegraphics[width=0.98\columnwidth]{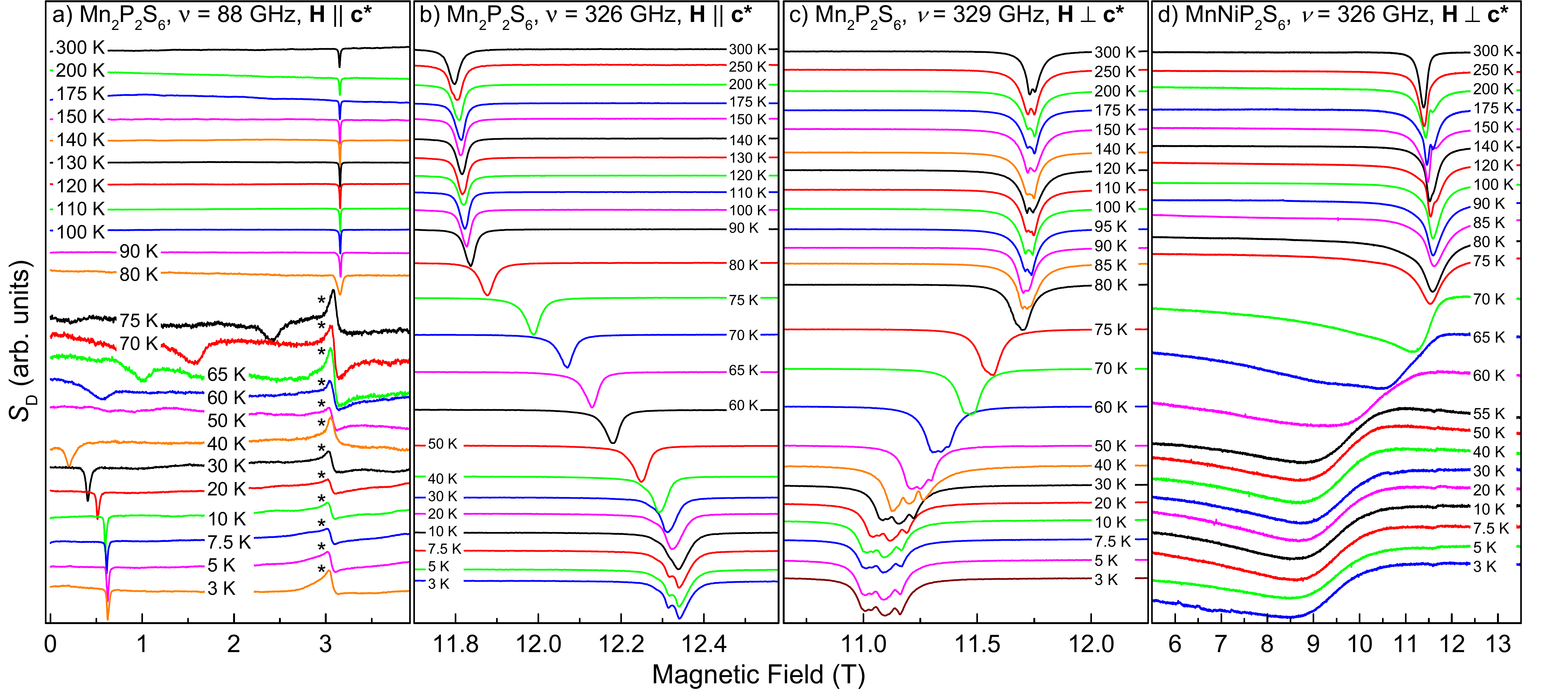}
	\caption{Temperature dependence of the HF-ESR spectra of (a) \MPS at $\nu \approx $ 88 GHz for \HIIc,  (b) \MPS at the excitation frequency, $\nu \approx $ 326 GHz for \HIIc configuration. The temperature independent peaks from the impurity in the probehead occurring only at low frequencies are marked with asterisks. (c) \MPS at $\nu \approx $ 329 GHz for \HIIab and (d) \MNPS at $\nu \approx $ 326 GHz for \HIIab. Spectra are normalized and vertically shifted for clarity.}
	\label{fig:Tdep_other}
\end{figure}

\twocolumngrid

\FloatBarrier

\bibliography{Reference}


\end{document}